\documentclass[onecolumn]{jpsj3} 

\def\runauthor{Online-Journal Subcommittee}

\title{%
Frustration-Induced Ferrimagnetism in Heisenberg Spin Chains \\
}

\author{%
Tokuro Shimokawa\thanks{E-mail address: rk09s002@stkt.u-hyogo.ac.jp }
 and Hiroki Nakano\thanks{E-mail address: hnakano@sci.u-hyogo.ac.jp}
}

\inst{%
Graduate School of Material Science, University of Hyogo, Kamigori, Hyogo 678-1297, Japan
}

\recdate{\today}

\abst{%
We study ground-state properties of the Heisenberg frustrated spin chain 
with interactions up to fourth nearest neighbors 
by the exact-diagonalization method and the density matrix renormalization group method. 
We find that ferrimagnetism is realized not only in the case of $S=$1/2 but also $S=$1 
despite that there is only a single spin site in each unit cell determined 
from the shape of the Hamiltonian. 
Our numerical results suggest that a ``multi-sublattice structure'' 
is not required for the occurrence of ferrimagnetism 
in quantum spin systems with isotropic interactions.}

\kword{%
quantum spin chain, frustration, ferrimagnetism, DMRG, exact diagonalization
}

\begin{document}
\maketitle

Ferrimagnetism is a fundamental phenomenon 
in the field of magnetism. 
One of the most typical examples of ferrimagnetism 
is the $(S, s)=(1, 1/2)$ mixed spin chain 
with a nearest-neighbor antiferromagnetic (AF) 
interaction\cite{Sakai}. 
In this system, 
the so-called Lieb-Mattis-type
ferrimagnetism\cite{Lieb, Marshall}  is realized in the ground state 
because two different spins 
are arranged alternately in a line owing to the AF interaction. 
This system includes two spins in a unit cell of the system.
In other known ferrimagnetic cases of quantum spin systems
except the $S=1/2$ Heisenberg frustrated spin chain studied in ref. \ref{Tokuro},
the situation that 
the system has more spins than one in each unit cell 
has been the same.  
Until our recent study\cite{Tokuro} demonstrated 
the occurrence of ferrimagnetism in the ground state 
of the $S=1/2$ Heisenberg frustrated spin chain 
despite the fact that a unit cell of the chain includes 
only a single spin, namely, it has no sublattice structure, 
it had been unclear whether the ``multi-sublattice structure'' is 
required for the occurrence of the ferrimagnetism 
in a quantum spin system composed of isotropic interactions.
The Hamiltonian examined in ref.~\ref{Tokuro} is given by
\begin{eqnarray}
\label{Hamiltonian}
\mathcal{H} &=& 
J \sum_{i}  [{\bf S}_{i}\cdot {\bf S}_{i+1}  
+  \mbox{$\frac{1}{2}$} {\bf S}_{i}\cdot {\bf S}_{i+2}] 
\nonumber \\ 
            & & - 
J^{\prime} \sum_{i}  [{\bf S}_{i}\cdot {\bf S}_{i+3} 
+ \mbox{$\frac{1}{2}$} ( {\bf S}_{i}\cdot {\bf S}_{i+2} 
+  A {\bf S}_{i}\cdot {\bf S}_{i+4}) ],
\end{eqnarray}
where the real constant $A$ is fixed to be unity. 
Here, ${\bf S}_{i}$ is the $S=1/2$ spin operator at the site $i$. 
The numerical study of this system clarified 
the existence of the ferrimagnetic ground state 
when the controllable parameter $J^{\prime}/J$ is changed. 
In addition, research confirmed that 
there are two types of ferrimagnetic phases: 
the phase of the Lieb-Mattis (LM) type and the phase of the non-Lieb-Mattis (NLM) type, 
which has been found in several frustrated spin 
systems\cite{PF4, PF5, Nakano-2D, strip}.

The purpose of this study is to confirm that 
the above example is not a special or rare case 
by investigating other models. 
In this study, we discuss the ground state 
of Hamiltonian (\ref{Hamiltonian}) 
not only in the case of $S=1/2$, 
but also in the case of 
${\bf S}_{i}$ being an $S=1$ spin operator.  
Moreover, we focus on the case of $A=0.4$, 
which is different from $A=1$. 
Note that energies are measured in units of $J$; 
we set $J=1$ hereafter. 

We employ two reliable numerical methods, i.e.,
the density matrix renormalization group (DMRG) method\cite{DMRG1,DMRG2} and 
the exact-diagonalization (ED) method. 
Both methods can give precise physical quantities 
for finite-size clusters. 
The DMRG method is very powerful for a one-dimensional system 
under the open-boundary condition. 
On the other hand, the ED method does not suffer 
from the limitation posed by the shape of the clusters; 
there is no limitation of boundary conditions,  
although the ED method can treat only systems smaller 
than those that the DMRG method can treat. 
Note that, in the present research, 
we use the ``finite-system'' DMRG method. 

\begin{figure}[t]
\begin{center}
\includegraphics[width=6.2cm]{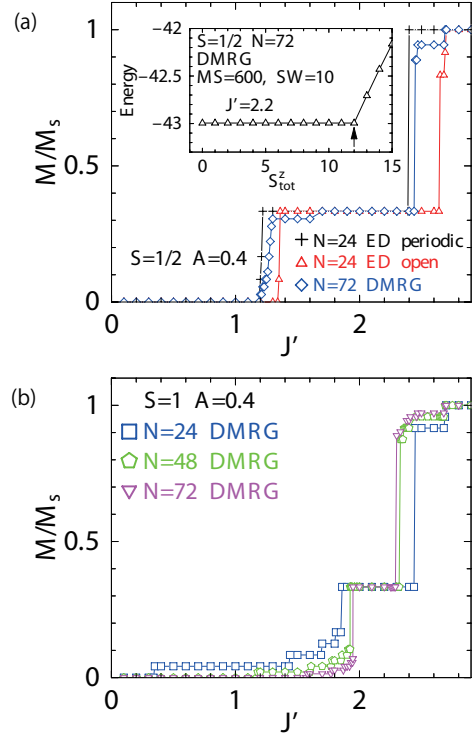}
\caption{(Color) (a) $J^{\prime}$ dependence 
of the normalized magnetization $M/M_{\rm s}$ 
in the ground state in the case of $S=1/2$ with $A=0.4$. 
In the inset of (a), the lowest energy in each subspace divided 
by $S_{\rm tot}^{z}$ is shown. 
Results of the DMRG calculations are presented 
when the system size is $N=72$ 
for $J^{\prime}=2.2$. 
The arrowhead indicates the spontaneous magnetization $M$ 
for a given $J^{\prime}$; $M$ is determined 
to be the highest $S_{\rm tot}^{z}$ 
among the values with the lowest common energy. 
(b) $J^{\prime}$ dependence 
of $M/M_{\rm s}$ 
in the ground state in the case of $S=1$ with $A=0.4$. 
}
\label{fig1}
\end{center}
\end{figure}

In the present study, 
two quantities are calculated. 
One is the lowest energy in each subspace divided 
by $S_{\rm tot}^{z}$ to determine 
the spontaneous magnetization $M$, 
where $S_{\rm tot}^{z}$ is the $z$ component of the total spin. 
We obtain the lowest energy $E(N,S_{\rm tot}^{z},J^{\prime})$ 
for a system size $N$ and a given $J^{\prime}$.  
For example, the $S_{\rm tot}^{z}$ dependence 
of $E(N,S_{\rm tot}^{z},J^{\prime})$ 
in a specific case of $J^{\prime}$ is presented 
in the inset of Fig. 1(a). 
This inset shows the results obtained by our DMRG calculations 
of the system of $N=72$ with the maximum number of retained states
 ($MS$) of 600, and a number of sweeps ($SW$) of 10. 
One can find the spontaneous magnetization $M$ 
for a given $J^{\prime}$ as the highest $S_{\rm tot}^{z}$ 
among those at the lowest common energy.  
(See the arrowhead in the inset.) 
The other quantity is the local magnetization in the ground state 
for investigating the spin structure 
of the highest-$S_{\rm tot}^{z}$ state. 
The local magnetization is obtained 
by calculating $\langle S_{i}^{z} \rangle$, 
where $S_{i}^{z}$ is the $z$-component 
of the spin at the site $i$ and 
$\langle O \rangle$ denotes the expectation value 
of the physical quantity $O$ 
with respect to the state of interest. 

\begin{figure}[h]
\begin{center}
\includegraphics[width=9.2cm]{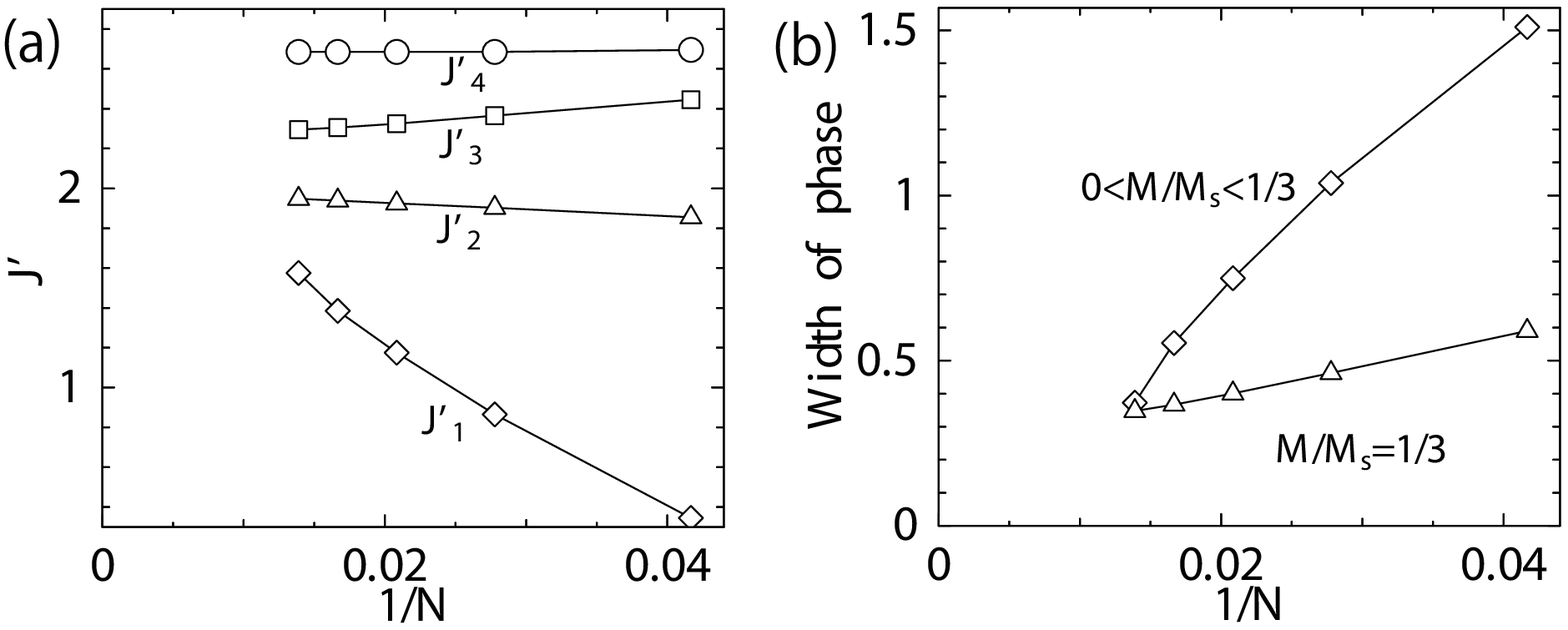}
\caption{(a) Size dependences of the boundaries 
of the regions in the case of $S=1$ with $A=0.4$. 
The results presented are those of $N=24, 36, 48, 60$, and 72 
from the DMRG calculations. 
(b) Size dependence of the width of each region in the case of $S=1$ with $A=0.4$.
The width of the region of $0<M/M_{\rm s}<1/3$ 
and that of $M/M_{\rm s}=1/3$ are defined 
as $|J_{2}^{\prime}-J_{1}^{\prime}|$ and  
$|J_{3}^{\prime}-J_{2}^{\prime}|$, respectively.  
}
\label{fig2}
\end{center}
\end{figure}

\begin{figure}[h]
\begin{center}
\includegraphics[width=6.5cm]{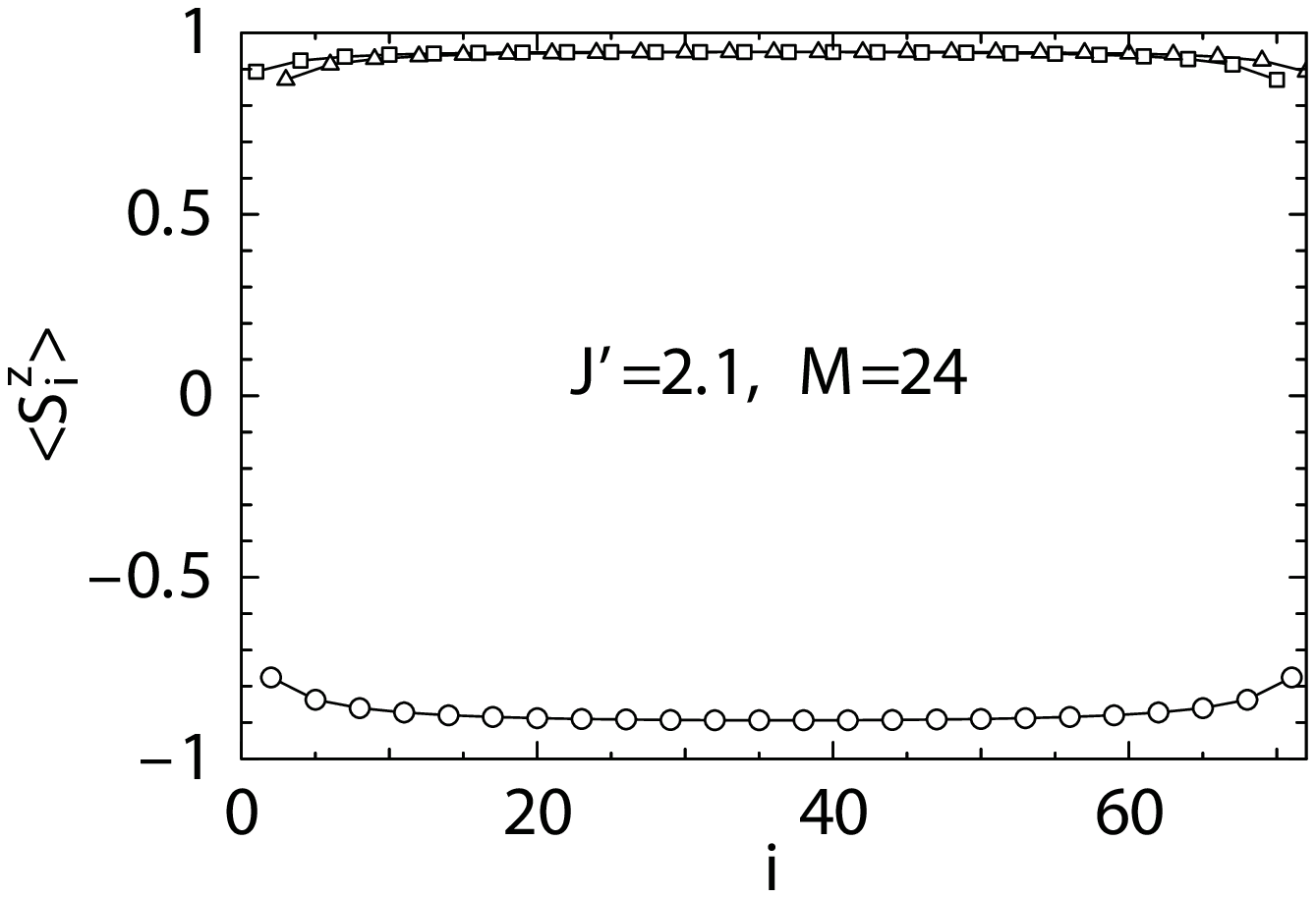}
\caption{Local magnetization $\langle S_{i}^{z} \rangle$ 
under the open-boundary condition: 
for $J^{\prime}=2.1$ in the case of $S=1$ with $A=0.4$
from the DMRG calculation for $N=72$. 
The site number is denoted by $i$, which is classified 
into $i=3n-2$, $3n-1$, and $3n$, where $n$ is an integer.
Squares, circles, and triangles mean $i=3n-2$, $3n-1$, and $3n$, 
respectively. }
\label{fig3}
\end{center}
\end{figure}

First, let us show the results of the $J^{\prime}$ dependence 
of $M/M_{\rm s}$ in Fig. \ref{fig1}, where $M_{\rm s}$ is 
the saturated magnetization.
Irrespective of $S=1/2$ or $S=1$, 
we find the nonmagnetic phase ($M/M_{\rm s}=0$) and 
ferromagnetic phase ($M/M_{\rm s}=1$). 
Between the two phases, we also find 
three regions: the regions of  $0<M/M_{\rm s}<1/3$, 
$M/M_{\rm s}=1/3$,  
and $1/3<M/M_{\rm s}<1$. 
For $S=1/2$, one can see that 
the region of $0<M/M_{\rm s}<1/3$ is much narrower than 
the distinctly existing region 
of NLM ferrimagnetism\cite{Tokuro} 
in the case of $S=1/2$ with $A=1$. 
The width of the present region for $A=0.4$ seems to vanish 
in the limit of $N \rightarrow \infty$. 
One finds that 
the occurrence of the NLM ferrimagnetism 
in Hamiltonian (\ref{Hamiltonian}) requires 
a fourth-neighbor interaction with $A$ that is larger than 
the specific value between $A=0.4$ and $A=1$. 
The width of the region 
of $M/M_{\rm s}=1/3$ 
in both cases of $S=1/2$ with $A=0.4$ and $S=1$ with $A=0.4$ 
seems to survive in the limit of $N \rightarrow \infty$. 
The region of $1/3<M/M_{\rm s}<1$ is presumably considered to 
merge with the ferromagnetic (FM) phase 
in the thermodynamic limit.
The reason for this is that this region appears only near $M/M_{\rm s}=1$ 
and that $M/M_{\rm s}$ in this region becomes progressively 
larger with increasing $N$. 
In addition, we cannot confirm this region 
in the calculations within $N\leq 30$ of the $S=1/2$ system 
under the periodic-boundary condition 
irrespective of the values of $A$. 
The issue of whether or not the region of $1/3<M/M_{\rm s}<1$ 
survives should be clarified in future studies; 
hereafter, we do not pay further attention to this issue. 

Next, we study the size dependences of the phase boundaries 
in the case of $S=1$ with $A=0.4$ depicted 
in Fig. \ref{fig2}(a). 
We present results of four boundaries:
$J^{\prime}=J_{1}^{\prime}$ 
between the nonmagnetic phase and 
the region of $0<M/M_{\rm s}<1/3$,  
$J^{\prime}=J_{2}^{\prime}$ 
between the regions of $0<M/M_{\rm s}<1/3$ and $M/M_{\rm s}=1/3$,  
$J^{\prime}=J_{3}^{\prime}$
between the regions of $M/M_{\rm s}=1/3$ and $1/3<M/M_{\rm s}<1$, and
$J^{\prime}=J_{4}^{\prime}$ 
between the region of $1/3<M/M_{\rm s}<1$ and the FM phase. 
To confirm the behavior up to the thermodynamic limit, 
we also examine the $N^{-1}$ dependences 
of the two widths of the regions of $M/M_{\rm s}=1/3$ and $0<M/M_{\rm s}<1/3$
in Fig. \ref{fig2}(b). 
Although the width of the region of $M/M_{\rm s}=1/3$ 
decreases with increasing $N$, 
this dependence shows a behavior that is convex-downwards 
for large sizes; the width seems to converge to 0.3.
Therefore, the phase of $M/M_{s}=1/3$ definitely survives 
in the limit of $N \rightarrow \infty$. 
On the other hand, 
the width of $0<M/M_{\rm s}<1/3$ obviously disappears 
in the limit of $N \rightarrow \infty$.
An appropriate tuning of the parameters 
in Hamiltonian (\ref{Hamiltonian}) of the $S=1$ system might 
cause the NLM ferrimagnetism; 
such parameter sets should be searched for in future studies. 

Finally, we examine the local magnetization 
$\langle S_{i}^{z} \rangle$ 
in the phase of $M/M_{\rm s}=1/3$ 
in the case of $S=1$ with $A=0.4$. 
In Fig. \ref{fig3}, we present our DMRG result 
of $\langle S_{i}^{z} \rangle$ of the system of $N=72$. 
We confirm the up-down-up spin behavior, 
and this spin structure is consistent with $M/M_{\rm s}$=1/3  
in the parameter region near approximately $J^{\prime}=2.1$ 
in Fig. \ref{fig1}(b).  
Thus, this phase is considered to be 
the LM-type ferrimagnetic phase.

In summary, 
we study the ground-state properties 
of a frustrated Heisenberg spin chain 
by the ED and DMRG methods.
Despite the fact that this system consists of 
only a single spin site in each unit cell determined 
from the shape of the Hamiltonian, 
the LM-type ferrimagnetic ground state is realized 
in a finite region not only in the case of $S=1/2$ 
but also of $S=1$. 
The present models showing ferrimagnetism indicate 
that a ``multi-sublattice structure'' is not 
required for the occurrence of ferrimagnetism 
in quantum spin systems with isotropic interactions 
as a general circumstance. 

We are grateful to Professor Y. Hasegawa for his critical reading of the manuscript.
This work was partly supported  
by Grants-in-Aid (Nos. 20340096, 23340109, and 23540388) 
from the Ministry of Education, Culture, Sports, 
Science and Technology (MEXT) of Japan.
This work was partly supported by 
a Grant-in-Aid (No. 22014012) 
for Scientific Research and Priority Areas 
``Novel States of Matter Induced by Frustration'' 
from the MEXT of Japan. 
Some of the calculations were carried out 
at the Supercomputer Center, 
Institute for Solid State Physics, University of Tokyo.
Exact-diagonalization calculations in the present work were 
carried out based on TITPACK Version 2 coded by H. Nishimori.
DMRG calculations were carried out 
using the ALPS DMRG application\cite{ALPS}.

\end{document}